\documentclass[12pt]{article}
\usepackage{amsfonts}
\usepackage{amssymb}
\usepackage{amsbsy}
\usepackage{mathrsfs}
\usepackage{amsmath}
\usepackage{graphicx}    % il parait que `graphicx' est requis pour `rotating'
\usepackage{rotating}    % necessaire pour avoir des gros tableaux tournes
\usepackage{epsfig}
\usepackage{color}
\usepackage{mathrsfs}

\input epsf.sty

\newlength{\TZ}
\newcommand{\BEQ}{\begin{equation}}     % Gleichungen Anfang ..
\newcommand{\BEA}{\begin{eqnarray}}
\newcommand{\BD}{\begin{displaymath}}
\newcommand{\EEQ}{\end{equation}}       % .. und Ende
\newcommand{\EEA}{\end{eqnarray}}
\newcommand{\ED}{\end{displaymath}}
\newcommand{\bb}{\begin{eqnarray}}
\newcommand{\ee}{\end{eqnarray}}

\newcommand{\vep}{\varepsilon}          % epsilon
\newcommand{\vph}{\varphi}              % rundes phi
\newcommand{\D}{{\rm d}}                % gerades d fuer Ableitungen
\newcommand{\II}{{\rm i}}               % gerades i fuer komplexe Einheit
\newcommand{\demi}{\frac{1}{2}}         % Bruch 1/2

\newcommand{\wit}[1]{\widetilde{#1}}    % weite Schlange
\renewcommand{\vec}[1]{\boldsymbol{#1}} % Vektoren fettgedruckt

\catcode`\@=11
\def\numberbysection{\@addtoreset{equation}{section}
        \def\theequation{\thesection.\arabic{equation}}}
\numberbysection

\begin{document}

\begin{titlepage}

\vskip 1.5 cm
\begin{center}
{\Large \bf From dynamical scaling to local scale-invariance: \\
a tutorial\footnote{Dedicated to Wolfhard Janke at the occasion of his 60$^{\rm th}$ birthday} 
}
\end{center}

\vskip 2.0 cm
\centerline{{\bf Malte Henkel}$^{a,b}$} 

\vskip 0.5 cm
\begin{center}
$^a$Rechnergest\"utzte Physik der Werkstoffe, Institut f\"ur Baustoffe (IfB), ETH Z\"urich, \\
Stefano-Franscini-Platz 3, CH - 8093 Z\"urich, Switzerland \\~\\
$^b$Groupe de Physique Statistique,
D\'epartement de Physique de la Mati\`ere et des Mat\'eriaux,
Institut Jean Lamour (CNRS UMR 7198), Universit\'e de Lorraine Nancy, \\
B.P. 70239,  F -- 54506 Vand{\oe}uvre l\`es Nancy Cedex, France\\~\\
\end{center}

\begin{abstract}
Dynamical scaling arises naturally in various many-body systems far from equilibrium. After a short historical overview, 
the elements of possible extensions of dynamical scaling to a local scale-invariance will be introduced. Schr\"odinger-invariance, 
the most simple example of local scale-invariance, will be introduced as a dynamical symmetry in the Edwards-Wilkinson 
universality class of interface growth. The Lie algebra construction, its representations and the Bargman superselection rules
will be combined with non-equilibrium Janssen-de Dominicis field-theory to produce explicit predictions for responses and
correlators, which can be compared to the results of explicit model studies. 

At the next level, the study of non-stationary states requires to go over, from Schr\"o\-din\-ger-in\-va\-riance, to ageing-invariance. 
The ageing algebra admits new representations, which acts as dynamical symmetries on more general equations, 
and imply that each non-equilibrium scaling operator is characterised by two distinct, 
independent scaling dimensions. Tests of ageing-invariance are described, in the Glauber-Ising and spherical models 
of a phase-ordering ferromagnet and the Arcetri model of interface growth.
%
%\centerline{\textcolor{red}{\Large \today}}
\end{abstract}

\vfill
%\noindent
%PACS numbers: 05.40.-a, 05.70.Ln, 81.10.Aj, 02.50.-r, 68.43.De \textcolor{blue}{\tt \`a revoir} 

\end{titlepage}

\setcounter{footnote}{0} 

%
%%%%%%%%%%%%%%%%%%%%%%%%%%%%%%%%%%%%%%%%%%%%%%%%%%%%%%%%%%%%%%%%%%%%%%%%%%%%%%%%%%%%%%%%%%%%%%%%%%%%%%%%%%%%%%%%%%%%
\section{Dynamical symmetries out of equilibrium} \label{intro}
%%%%%%%%%%%%%%%%%%%%%%%%%%%%%%%%%%%%%%%%%%%%%%%%%%%%%%%%%%%%%%%%%%%%%%%%%%%%%%%%%%%%%%%%%%%%%%%%%%%%%%%%%%%%%%%%%%%%
%
Symmetries have played an important r\^ole in physics since a long time \cite{Weyl52}, and new types of symmetry and new applications are 
continuously being discovered. The best-known example of a time-space symmetry is the special-relativistic Poincar\'e-invariance, 
of either classical mechanics or classical electrodynamics \cite{Einstein05}. While the principle of relativity acts essentially as
a scaffold on which more specific physical theories can be constructed, larger time-space symmetries can be realised if the
physical system under study is specified in more detail: for example, Maxwell's equations of a free electromagnetic field in the
vacuum, in $d=1+3$ time-space dimensions,\footnote{Remarkably, the conformal invariance of the free Maxwell field no longer holds true
in $d\ne 4$ time-space dimensions, although the theory certainly is scale-invariant: 
for $d=3$, the theory can be embedded into an unitary conformally invariant
field-theory, while for $d\geq 5$, only non-unitary extensions exist \cite{Showk11}.} 
admit a conformal symmetry \cite{CuBa0910}. Continuous phase transitions, 
at thermodynamic equilibrium, constitute a very widespread set of examples 
where the strong interactions of a large number of degrees of freedom may create first a scale-invariance \cite{StuWidKad}, at least at certain
specific `critical points' in parameter space,\footnote{For historical overviews on critical phenomena see \cite{Fisher98}, and also \cite{Berche09}.} 
which in many `favourable' cases can be extended further to conformal invariance \cite{Polyakov70,Belavin84}. In particular, there is
a proof of conformal invariance in  the Ising model universality class, in any dimension \cite{Delamotte16}. 
Schematically, scale-invariance defines the critical exponents and through the renormalisation group establishes mainly qualitative properties,
such as their universality (and produces the scaling relations between the exponents) but does not fix their values. Conformal invariance
rather makes quantitative predictions in fixing the form of the scaling functions and, at least in $d=2$ dimensions, produces the admissible
values of the exponents from the unitary representations of the Virasoro algebra. 
At a phase transition, conformal invariance is a property of the effective theory, which describes the long-distance properties of a critical
system. On the other hand, conformal invariance also arises as a `fundamental' symmetry from the reparametrisation-invariance in string theory, 
see \cite{PolZwie} and references therein.  

In condensed-matter and non-equilibrium statistical physics, one is often led to study time-dependent critical phenomena, of which
Brownian motion is one of the best-known examples \cite{EinLang}. Their time-space dynamical symmetries have since a very long time been
known to mathematicians \cite{Lie1881,Jacobi1842} as a dynamical symmetry, originally either of the motion of free particles or of free diffusion of
an ensemble of particles, and the corresponding Lie group is nowadays usually called the {\em Schr\"odinger group}. 
This Lie group, and it associated Lie algebra, caught the attention of physicists much later \cite{Niederer72,Hagen72,Burdet72,Jackiw72}. 
Here, we shall be interested in a class of applications of extensions of dynamical scaling to the collective non-equilibrium behaviour, as
it arises (i) in the phase-ordering kinetics of simple magnets \cite{Bray94} quenched into the coexistence phase below the
critical temperature $T_c>0$, from a disordered initial state, or (ii) in the kinetics of interface growth \cite{Barabasi95,Halpin95,Krug97}. 
The description of these examples of non-equilibrium critical phenomena
owes a lot to earlier studies on the {\em physical ageing} in glassy and non-glassy systems \cite{Cugliandolo02,Henkel10,Taeuber14}. 
Remarkably, experiments on the mechanical relaxation in many polymer systems, to be followed later by analogous studies in many different kinds of 
glassy and non-glassy systems, established that physical ageing has indeed many {\em reproducible} and {\em universal} aspects \cite{Struik78}. 
This allows one to present a formal definition of ageing in complex physical systems \cite{Henkel10}: \\

\noindent 
{\em a system undergoes {\em physical ageing}\footnote{We use european spelling throughout.} if its relaxational
behaviour has the following properties:
%\newpage
\begin{enumerate}
\item slow relaxation of the observables (not described in terms of a single exponential)
\item breaking of time-translation-invariance
\item dynamical scaling \footnote{Physically, this means that there is a single time-dependent length scale $L(t)$, but {\it a priori} nothing is
yet said on its precise form. Indeed, for systems with frustrations and/or disorder, one expects for large times logarithmic growth
$L(t)\sim \ln^{1/\psi} t$ (eventually after a very long cross-over regime), 
whereas for simple systems without disorder and frustrations, an algebraic law $L(t)\sim t^{1/z}$ is expected. We
shall restrict throughout to the latter case, also referred to as {\em simple ageing}.} 
\end{enumerate} }

A central quantity for the description of such non-equilibrium systems is either the time-space-dependent magnetisation $m(t,\vec{r})$ in the case
of phase-ordering or the interface height $h(t,\vec{r})$ for interface growth. Both are instances of a time-space-dependent 
{\em order-parameter} $\vph=\vph(t,\vec{r})$. Adopting a continuum description, 
this order-parameter is assumed to obey a stochastic Langevin equation
\BEQ \label{1}
2{\cal M}\partial_t \vph= \Delta_{\vec{r}} \vph - \frac{\delta\mathscr{V}[\vph]}{\delta \vph} + \left(\frac{T}{\cal M}\right)^{1/2} \eta
\EEQ 
where $\cal M$ is a kinetic coefficient, $\Delta_{\vec{r}}$ the spatial laplacian 
and the potential $\mathscr{V}[\vph]$ fixes the detailed behaviour of the model. The solution $\vph$ is
a random variable, since $\eta$ is a centred gaussian noise of unit variance and $T$ plays the r\^ole of a temperature and also, the initial state is
assumed to obey a centred gaussian distribution with variance $\langle \vph(0,\vec{r})\vph(0,\vec{r}')\rangle=\Delta_0 \delta(\vec{r}-\vec{r}')$. 
The special case $T=0$ includes the physics of phase-ordering kinetics, while the special case $\Delta_0=0$ includes interface growth. 
In both cases, one has $\langle \vph(t,\vec{r})\rangle=0$, where $\langle\cdot\rangle$ denotes the average over the thermal or initial distributions. 
See table~\ref{tab1} for a schematic comparison.\footnote{For magnets, a quench to $T=T_c$ produces critical dynamics: depending on the initial state
either {\it at} or {\it out of } equilibrium.} 

%%%==========================================================================================
\begin{table}[tb]
\caption[tab1]{Comparison between the kinetics of phase-ordering and of interface growth \label{tab1}}
\begin{tabular}{ll}      \hline\noalign{\smallskip}
phase-ordering                                       & interface growth \\  \noalign{\smallskip}\hline\noalign{\smallskip}
thermodynamic equilibrium state                      & growth continues forever \\
magnetisation $m(t,\vec{r})$                         & height profile $h(t,\vec{r})$ \\
phase transition at $T=T_c$                          & same generic behaviour for $T>0$ \\
~~(ageing for $T\leq T_c$, no ageing for $T>T_c$)    & ~~(deterministic for $T=0$, probabilistic for $T>0$)\\
variance $\left\langle \left( m(t,\vec{r})-\langle m(t)\rangle\right)^2 \right\rangle\sim t^{-\beta/(z \nu)}$ &
roughness $\left\langle \left( h(t,\vec{r})-\langle h(t)\rangle\right)^2 \right\rangle\sim t^{\beta}$ \\
relaxation, after quench to $T\leq T_c$              & relaxation, from initial substrate \\
autocorrelator $C(t,s) = \langle m(t) m(s)\rangle_c$ & autocorrelator $C(t,s)=\langle h(t) h(s)\rangle_c$  \\
                         \noalign{\smallskip}\hline
\end{tabular}
\end{table}
%%%==========================================================================================
 
\begin{figure}[tb]
\resizebox{0.65\columnwidth}{!}{%
  \includegraphics{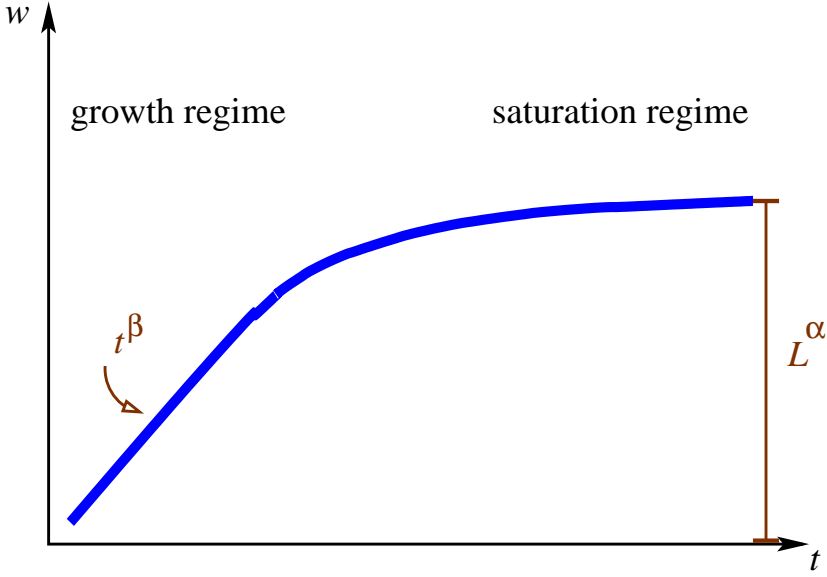} }
\caption[fig1]{Schematic evolution of the interface width on a substrate with linear dimension $L$. The growth regime (where ageing occurs) 
and the saturation regime, both with the associated scaling, are indicated. \label{fig1}}       
\end{figure}

A large part of this work will concentrate on the paradigmatic special case $\mathscr{V}=0$. 
As we shall see, it already contains many important features
of more general systems which can thereby explained in a simple way. 
In our  paradigm, the case $T=0$ is often called the {\em free gaussian model} of
phase-ordering kinetics, whereas the case $\Delta_0=0$ is known as the {\em Edwards-Wilkinson model} of interface growth \cite{Edwards82}. 
In the latter situation, the interface width, on a hypercubic lattice $\mathscr{L}\subset\mathbb{Z}^d$ with $|\mathscr{L}|=L^d$ sites, 
usually shows, for large times, {\em Family-Vicsek scaling} \cite{Family85}
\BD 
\!w^2(t;L) := \frac{1}{L^d} \sum_{\vec{r}\in\mathscr{L}} \left\langle \left( h(t,\vec{r})-\overline{h}(t) \right) \right\rangle^2 
= L^{2\beta z} f_w(t L^{-z}) 
\sim \left\{ \begin{array}{ll} t^{2\beta} & \mbox{\rm ~;~ if $tL^{-z}\ll 1$} \\
                              L^{2\alpha} & \mbox{\rm ~;~ if $tL^{-z}\gg 1$} \end{array} \right.
\ED
where $\overline{h}(t)$ is the spatially averaged height, $\beta$ is the {\em growth exponent} 
and $\alpha$ is the {\em roughness exponent}, see figure~\ref{fig1}. The dynamical exponent $z=\alpha/\beta>0$.  
When $t L^{-z}\gg 1$, one speaks of the {\em saturation regime} and when $t L^{-z}\ll 1$, one speaks of the {\em growth regime}. 
All studies of ageing in interface growth are in the growth regime, on which we shall focus from now on. 
In contrast to equilibrium critical phenomena, non-equilibrium scaling, as in phase-ordering or interface growth, can be achieved
{\em without} having to fine-tune one or several thermodynamic parameters of the macroscopic system. 

%%%++++++++++++++++++++++++++++++++++++++++++++++++++++++++++++++++++++++++++++++++++++++++++
%%%++++++++++++++++++++++++++++++++++++++++++++++++++++++++++++++++++++++++++++++++++++++++++

The ageing behaviour of the solutions of the Langevin equation (\ref{1}) 
is conveniently studied through the two-time correlators $C$ and responses $R$, defined as
\begin{subequations} \label{2}
\begin{align}
C(t,s;\vec{r}) &= \langle \vph(t,\vec{r}+\vec{r}_0) \vph(s,\vec{r}_0) \rangle 
                - \langle \vph(t,\vec{r}+\vec{r}_0)\rangle\langle \vph(s,\vec{r}_0)\rangle
\label{2C} \\
R(t,s;\vec{r}) &= \left. \frac{\delta \langle \vph(t,\vec{r}+\vec{r}_0)\rangle}{\delta j(s,\vec{r}_0)}\right|_{j=0} 
= \langle \vph(t,\vec{r}+\vec{r}_0) \wit{\vph}(s,\vec{r}_0) \rangle
\label{2R}
\end{align}
\end{subequations}
where $j=j(t,\vec{r})$ is an external perturbation
conjugate to the order-parameter $\vph$, to be added to eq.~(\ref{1}), and $\wit{\vph}$ is the associated response field, 
in the context of Janssen-de Dominicis field theory, see \cite{Taeuber14}. 
In many cases, for instance when $\mathscr{V}=0$, spatial translation-invariance holds true, as anticipated in (\ref{2}). 
In addition, one has the following dynamical scaling (also assuming rotation-invariance for $d>1$ dimensions)
\BEQ \label{3}
C(t,s;\vec{r}) = s^{-b} F_C\left(\frac{t}{s};\frac{|\vec{r}|^z}{t-s}\right) \;\; , \;\;
R(t,s;\vec{r}) = s^{-1-a} F_R\left(\frac{t}{s};\frac{|\vec{r}|^z}{t-s}\right)
\EEQ
which defines the ageing exponents $a,b$. The scaling forms (\ref{3}), often referred to as {\em simple ageing}, implicitly assume the
existence of a single time-dependent length scale $L=L(t)$, and with an algebraic long-time behaviour $L(t)\sim t^{1/z}$, which
defines the {\em dynamical exponent} $z$. Often, one focuses on the autocorrelator and the autoresponse
\begin{subequations} \label{4}
\begin{align}
C(t,s) &= C(t,s;\vec{0}) = s^{-b}   f_C\left(\frac{t}{s}\right) \hspace{0.42truecm}\;\; , \;\; 
                  f_C(y) = F_C\left(y;0\right) \stackrel{y\gg 1}{\sim} y^{-\lambda_C/z} 
\label{4C} \\
R(t,s) &= R(t,s;\vec{0}) = s^{-1-a} f_R\left(\frac{t}{s}\right) \;\; , \;\; 
                  f_R(y) = F_R\left(y;0\right) \stackrel{y\gg 1}{\sim} y^{-\lambda_R/z} 
\label{4R}
\end{align}
\end{subequations}
and defines the {\em autocorrelation exponent} $\lambda_C$ and the {\em autoresponse exponent} $\lambda_R$. 
The exponent $b$ is simply related to stationary exponents: one has $b=0$ in phase-ordering, $b=2\beta/(z\nu)$ for critical dynamics
and $b=-2\beta$ for interface growth. The value of $a$ can be fixed if a fluctuation-dissipation theorem ({\sc fdr}), i.e. a relationship between
$C$ and $R$, holds true.\footnote{At equilibrium, one has Kubo's well-known result: $T R(t-s;\vec{r})=\partial_s C(t-s;\vec{r})$. Non-equilibrium
stationary systems with a known {\sc fdr} include critical directed percolation \cite{Enss04,Baumann07} or the Kardar-Parisi-Zhang universality class 
in one dimension \cite{Lvov93,Frey96}. Their {\sc fdr}s are distinct from Kubo's form.} 
In the known cases where such a relationship exists, this also implies $\lambda_C=\lambda_R$, but for a fully
disordered initial state, the autocorrelation and autoresponse exponents are independent of all equilibrium exponents, 
see \cite{Henkel10,Taeuber14}.\footnote{Since response functions, defined in (\ref{2R}), are difficult to measure directly in a numerical simulation,
it is common practise to use time-integrated dynamical susceptibilities instead, as introduced first in glassy systems \cite{Cugliandolo02}. 
However, there are several pitfalls for the correct interpretation of the scaling of dynamical susceptibilities, especially so for phase-ordering
\cite{Henkel0304}, see \cite{Henkel10} for full details. For interfaces, the analogues of dynamical susceptibilities are computed from a
damage-spreading simulation \cite{Henkel12}.}  

{\em Does there exist any extension of dynamical scaling, which would constrain the form of the scaling functions in (\ref{4})~? 
Can one use conformal invariance, at equilibrium critical points, as a guide to find such extensions~?} 

In order to
present the basic elements of a possible answer,\footnote{Presented for the first time in \cite{Henkel92}.} 
we shall concentrate for quite a while on the paradigmatic case $\mathscr{V}[\vph]=0$ in eq.~(\ref{1}). This gives the 
{\em Edwards-Wilkinson equation} \cite{Edwards82}. It
is well-known that its dynamical exponent $z=2$. For $z=2$, an extension of dynamical scaling is given by the {\em Schr\"odinger group} of
time-space transformations
\BEQ \label{5}
t \mapsto t'=\frac{\alpha t+\beta}{\gamma t +\delta} \;\; , \;\;
\vec{r} \mapsto \vec{r}'=\frac{\mathscr{R}\vec{r}+\vec{v}t+\vec{a}}{\gamma t+\delta} \;\; ; \;\; \alpha\delta-\beta\gamma=1
\EEQ
with a rotation matrix $\mathscr{R}\in{\sl SO}(d)$, $\vec{v},\vec{a}\in\mathbb{R}^d$ and $\alpha,\beta,\gamma,\delta\in\mathbb{R}$. 
The transformation in time is indeed a (projective) conformal transformation, the transformations in space are rotations, Galilei-transformations
and translations, as parametrised by $\mathscr{R}$, $\vec{v}$ and $\vec{a}$. The Schr\"odinger group is {\em not} semi-simple, its representations
are therefore projective: co-variant solutions of Schr\"odinger-invariant equations transform also through the presence of a 
`companion function'.\footnote{This is well-known from the Galilei-transformation of the wave function $\psi(t,\vec{r})$ of a free particle
in non-relativistic quantum mechanics.} 
Consequently, since the Edwards-Wilkinson equation, eq.~(\ref{1}) with $\mathscr{V}[\vph]=0$, 
describes the coupling  of the system with a bath, this coupling is incompatible with any dynamical symmetries, 
beyond translation- and rotation-invariance, and dynamical scaling. Therefore, symmetries of (\ref{1}) will be studied in two steps, as follows:
\begin{enumerate}
\item find the dynamical symmetries of the noiseless simple diffusion equations, with $T=0$ and $\Delta_0=0$ \cite{Lie1881,Niederer72}. 
In particular, derive the Bargman superselection rules \cite{Bargman54} 
which follow from the combination of spatial translation-invariance and Galilei-invariance. 
\item using the non-equilibrium Janssen-de Dominicis theory, derive reduction formul{\ae}, in order to express any correlator or response
of the full, noisy theory in terms of averages computed only in  terms of the noise-less, deterministic theory \cite{Picone04}. 
\end{enumerate}
Therefore, Schr\"odinger-invariance of a Langevin equation (\ref{1}) is a {\em hidden symmetry} in the sense that formally it is only a symmetry of
its {\em deterministic part}, obtained from (\ref{1}) by setting $T=0$ and $\Delta_0=0$. 

The quest for {\em local scale-invariance} ({\sc lsi}) is to find non-trivial extensions of dynamical scaling which would allow
(i) to predict the form of the scaling functions of responses and correlators, once the scaling dimensions are known and (ii)
to fix, or at least to constrain, the values of these scaling dimensions. At present, some progress has been achieved on the first objective, 
while the second is still out of reach. Schr\"odinger-invariance is the most simple example of {\sc lsi}. It arises as dynamical symmetry of
the Edwards-Wilkinson equation and will be discussed at length
in section~2. Section~3 considers what may happen for truly non-equilibrium systems where time-translation-invariance no longer holds. 
Then we must instead study the {\em ageing algebra} $\mathfrak{age}(d)$, a true subalgebra of the Schr\"odinger algebra $\mathfrak{sch}(d)$. 
New features arise in the representations of $\mathfrak{age}(d)$ and we shall present some of the physical consequences.
The $1D$ Glauber-Ising model, the spherical model of a ferromagnet and the Arcetri model of interface growth are examples of ageing-invariant systems. 

We close with a short overview of some further tests and a brief outlook on current and possible future work. 
This tutorial is not intended as a review: we did not attempt completeness of neither the themes treated, nor the references quoted.

%
%%%%%%%%%%%%%%%%%%%%%%%%%%%%%%%%%%%%%%%%%%%%%%%%%%%%%%%%%%%%%%%%%%%%%%%%%%%%%%%%%%%%%%%%%%%%%%%%%%%%%%%%%%%%%%%%%%%%
\section{Schr\"odinger-invariance and the \\ Edwards-Wilkinson equation} \label{sect2}
%%%%%%%%%%%%%%%%%%%%%%%%%%%%%%%%%%%%%%%%%%%%%%%%%%%%%%%%%%%%%%%%%%%%%%%%%%%%%%%%%%%%%%%%%%%%%%%%%%%%%%%%%%%%%%%%%%%%
%
Here the elements of the dynamical symmetry of the Edwards-Wilkinson equation will be presented one after the other, step by step. 
We shall show how to derive the form of the Schr\"odinger-covariant correlators and responses. 

\subsection{Schr\"odinger algebra}

It is convenient to consider the infinitesimal form of the Schr\"odinger transformations (\ref{5}). 
The corresponding infinitesimal generators are, for technical simplicity in $d=1$ dimensions
\begin{subequations} \label{6} 
\begin{align}
X_n &= -t^{n+1}\partial_t - \frac{n+1}{2} t^n r\partial_r - \frac{x}{2}(n+1) t^n - \frac{n(n+1)}{4}{\cal M} t^{n-1} r^2 
\label{6X} \\
Y_m &= -t^{m+1/2}\partial_r - \left(m+\demi\right){\cal M} t^{m-1/2}r \label{6Y} \\
M_n &= -t^n {\cal M} \label{6M}
\end{align}
\end{subequations}
where $n\in\mathbb{Z}$ and $m\in\mathbb{Z}+\demi$. The Lie algebra of the finite-dimensional Schr\"odinger group in (\ref{5}) is
the {\em Schr\"odinger algebra} $\mathfrak{sch}(1) = \mbox{\rm Lie {\sl Sch}}(1) = \left\langle X_{\pm 1,0}, Y_{\pm 1/2}, M_0\right\rangle$,
see table~\ref{tab2} for their interpretation. 
%%%==========================================================================================
\begin{table}[tb]
\caption[tab2]{Lie algebra generators $\mathscr{X}$ of the Schr\"odinger algebra $\mathfrak{sch}(1)$. \label{tab2}}
\begin{tabular}{lll}  \hline\noalign{\smallskip}
$\mathscr{X}$ & generator  & interpretation \\
\noalign{\smallskip}\hline\noalign{\smallskip}
$X_{-1}$ & $-\partial_t$ & time-translation \\
$Y_{-1/2}$ & $-\partial_r$ & space-translation \\
$M_0$ & $-{\cal M}$ & phase shift \\
$X_0$ & $-t\partial_t - \demi r\partial_r - \frac{x}{2}$ & dilatation \\
$Y_{1/2}$ & $-t\partial_r - {\cal M} r$ & Galilei-transformation \\
$X_1$ & $-t^2\partial_t - tr\partial_r - x t - \demi {\cal M} r^2$ & `special' Schr\"odinger transformation \\
\noalign{\smallskip}\hline
\end{tabular}
\end{table}
%%%==========================================================================================
Then the non-vanishing commutators are
\BEA
\left[ X_n, X_{n'}\right] &=& (n-n') X_{n+n'} \;\; , \;\; 
\left[ X_n, Y_m\right] = \left(\frac{n}{2}-m\right) Y_{n+m} \nonumber \\
\left[ X_n, M_{n'} \right] &=& -n' M_{n+n'} \hspace{0.56truecm}\;\; , \;\; \left[ Y_{m}, Y_{m'} \right] = (m-m') M_{m+m'}
\label{7}
\EEA
This shows the inclusion $\mathfrak{sch}(1)\subset\mathfrak{sv}(1)$ of the six-dimensional Schr\"odinger algebra into the
infinite-dimensional {\em Schr\"odinger-Virasoro algebra} $\mathfrak{sv}(1)=\langle X_n, Y_{n+1/2}, M_n\rangle_{n\in\mathbb{Z}}$ 
\cite{Henkel92,Henkel94}. Integrating these infinitesimal transformations gives the Schr\"odinger-Virasoro Lie group
$t\mapsto t'$, $\vec{r}\mapsto \vec{r}'$ and $\vph\mapsto \vph'$ \cite{Henkel03,Unterberger12}. From the $X_n$, one has,  
\BEQ \label{8}
t = \beta(t') \;\; , \;\; \vec{r} = \vec{r}' \dot{\beta}(t')^{1/2}  \;\; , \;\;
\vph(t,\vec{r}) = \dot{\beta}(t')^{-x/2} \exp\left[ -\frac{{\cal M}{\vec{r}'}^2}{4} \frac{\dot{\beta}(t')}{\beta(t')} \right]
\vph'(t',\vec{r}')
\EEQ
with $\dot{\beta}(t)=\D\beta(t)/\D t$ and $\beta(t)$ is an arbitrary, but non-decreasing, 
differentiable function of time. Herein, $x$ and $\cal M$, respectively, are the scaling dimension and the mass of $\vph$. From the $Y_m$, one has 
\BEQ \label{8Y}
t=t' \;\; , \;\; \vec{r} = \vec{r}'-\vec{\alpha}(t') \;\; , \;\;
\vph(t,\vec{r}) = \exp\left[ {\cal M} \left(\demi \dot{\vec{\alpha}}(t')\cdot\vec{\alpha}(t') - \vec{r}'\cdot\vec{\alpha}(t')\right)\right]
\vph'(t',\vec{r}')
\EEQ
with a differentiable vector function $\vec{\alpha}(t)$ of time. From the $M_n$,  one obtains a time-dependent phase-shift. 
In contrast to the conformal group, this infinite-dimensional extension is possible for all dimensions $d\geq 1$, with an obvious
extension of the commutators (\ref{7}) \cite{Henkel10}. 

\subsection{Schr\"odinger-invariance of the diffusion equation}

The deterministic part of the Edwards-Wilkinson equation reads ${\cal S}\vph=0$, with the Schr\"odinger operator 
${\cal S}=2{\cal M}\partial_t - \Delta_{\vec{r}}$. The Schr\"odinger operator commutes with several elements of $\mathfrak{sch}(1)$:
\BEQ \label{10}
\left[ {\cal S},X_{-1}\right] = \left[{\cal S},Y_{\pm 1/2}\right] = \left[ {\cal S}, M_0 \right]=0
\EEQ
such that the corresponding transformations are symmetries of the operator $\cal S$, which takes here a r\^ole analogously to the
hamiltonian with respect to symmetries in quantum mechanics. However, there are two non-vanishing commutators
\BEQ \label{11}
\left[ {\cal S}, X_0 \right] = -{\cal S} \;\; , \;\;
\left[ {\cal S}, X_1 \right] = -2t{\cal S} +(2x-1)
\EEQ
If one considers a solution $\vph_0$ of the equation ${\cal S}\vph_0=0$, then $\mathscr{X}\vph_0$ is the transformed solution, 
with $\mathscr{X}=X_{0,1}\in\mathfrak{sch}(1)$. Now, (\ref{11}) implies that also 
${\cal S}(\mathscr{X}\vph_0)=0$, but only if the scaling dimension $x=\demi$. 
We consider {\it `symmetries'}\  in a generalised sense: 
a {\em dynamical symmetry} is the Lie algebra $\mathfrak{g}$ of transformations $\vph_0\mapsto \mathscr{X}\vph_0$, 
with $\mathscr{X}\in\mathfrak{g}$, which leave the solution space of ${\cal S}\vph_0=0$ invariant.  

Generalising to $d$ dimensions, one has: {\it the free diffusion equation ${\cal S}\vph=0$ is Schr\"odinger-invariant, 
i.e. its space of solutions is invariant under the action of $\mathfrak{sch}(d)$, 
if the scaling dimension $x=x_{\vph}=d/2$} \cite{Lie1881,Niederer72}. 

\subsection{Ward identities}

Consider $n$-point correlation functions 
\BD
C^{(n)} = C^{(n)}(t_1,\ldots,t_n;\vec{r}_1,\ldots,\vec{r}_n)=\langle \vph_1(t_1,\vec{r}_1)\ldots\vph_n(t_n,\vec{r}_n)\rangle
\ED
built from scaling operators $\vph_i$. Such a $C^{(n)}$ is {\em $\mathfrak{sch}$-covariant}, if it vanishes under the action of its infinitesimal
generators $\mathscr{X}^{[n]} C^{(n)}=0$ with $\mathscr{X}^{[n]}=\sum_{i=1}^n \mathscr{X}_i$ and $\mathscr{X}_i$ can be any of the generators
$\mathscr{X}\in\mathfrak{sch}(d)$ acting on the $i^{\rm th}$ scaling operator $\vph_i$.\footnote{Analogous to correlators of
quasi-primary scaling operators in conformal field-theory \cite{Polyakov70}.} These {\em Ward identities} permit to restrict the form of the $C^{(n)}$. 

\subsection{Bargman superselection rule}

As an example, we consider the consequences of spatial translation- and Galilei-invariance. 
The $n$-body operators are, with $D_i=\partial_{r_i}$ and for $d=1$
\BEQ
Y_{-1/2}^{[n]} = \sum_{i=1}^n \left[ - D_i \right] \;\; , \;\; 
Y_{1/2}^{[n]} = \sum_{i=1}^n \left[ -t_i D_i - {\cal M}_i r_i \right]
\EEQ
The Ward-identities $Y_{-1/2}^{[n]} C^{(n)} =Y_{-1/2}^{[n]} C^{(n)} =0$ lead to the differential equations
\begin{subequations} \label{12}
\begin{align}
\sum_{i=1}^n \frac{\partial}{\partial r_i}\, C^{(n)}(t_1,\ldots,t_n;r_1,\ldots,r_n) &= 0 \label{12Tra} \\
\sum_{i=1}^n \left[ t_i \frac{\partial}{\partial r_i} + {\cal M}_i r_i \right] C^{(n)}(t_1,\ldots,t_n;r_1,\ldots,r_n) &= 0 \label{12Gal}
\end{align}
\end{subequations}
Eq.~(\ref{12Tra}) implies that $C^{(n)}=C^{(n)}(t_1,\ldots,t_n;u_1,\ldots,u_{n-1})$, with $u_i=r_i-r_n$, which we abbreviate as
$C^{(n)}(\{t\};\{u\})$. Then eq.~(\ref{12Gal}) becomes
\BD
\sum_{i=1}^{n-1} \left[ -(t_i-t_n) \frac{\partial}{\partial u_i} - {\cal M}_i u_i \right] C^{(n)}(\{t\};\{u\}) 
+ r_n \left( {\cal M}_1 +\ldots +{\cal M}_n\right) C^{(n)}(\{t\};\{u\})   =0.
\ED
Because of spatial translation-invariance, an explicit dependence on $r_n$ is inadmissible. Hence, the last term must vanish, leading to
the {\em Bargman superselection rule} \cite{Bargman54} 
\BEQ \label{14}
\left( {\cal M}_1 +\ldots +{\cal M}_n\right) C^{(n)}(\{t\};\{u\})   =0
\EEQ

\subsection{Non-equilibrium field-theory}

The Langevin equation (\ref{1}) can be recast as the equation of motion following from a dynamical functional. Formally, the essential steps are
as follows, see \cite{Taeuber14} for details. Consider the average of an observable $\mathscr{A}$ 
\BEQ
\langle \mathscr{A}\rangle = \int \mathscr{D}\eta\: {\cal P}[\eta] \int \mathscr{D}\vph\;  \mathscr{A}[\vph]\: 
\delta\left( \left(2{\cal M}\partial_t - \Delta_{\vec{r}}\right)\vph - \mathscr{V}'[\vph] -j\vph -\sqrt{\frac{T}{\cal M}\,}\, \eta \right)
\EEQ
Here, ${\cal P}[\eta]$ is the distribution of the noise $\eta$, assumed gaussian. This noise represents the average over the initial conditions
for phase-ordering and is `thermal' for interface growth. One also uses an integral representation of the Dirac distribution
$\delta(x) = (2\pi)^{-1}\int_{\mathbb{R}} \!\D\wit{\vph}\: \exp(\II \wit{\vph} x)$. 
Inserting this into the generating functional, the gaussian integrals
over the noises can be carried out and one finally arrives at 
\BEQ \label{16}
\langle \mathscr{A}\rangle = \int \mathscr{D}\wit{\vph} \mathscr{D}\vph\; \mathscr{A}[\vph]\: \exp(-{\cal J}[\vph,\wit{\vph}\,])
\EEQ
where the dynamical functional ${\cal J}[\vph,\wit{\vph}\,]={\cal J}_0[\vph,\wit{\vph}\,]+{\cal J}_b[\wit{\vph}\,]$ is naturally decomposed into a 
{\em `deterministic part'} ${\cal J}_0$ and a {\em `noise part'} ${\cal J}_b$.These take the form
\BEQ \label{17}
{\cal J}_0[\vph,\wit{\vph}\,] = \int\!\D t\D \vec{r}\: 
\wit{\vph}\left( \left(2{\cal M}\partial_t -\Delta_{\vec{r}} -j\right)\vph- \mathscr{V}'[\vph]  \right)
\; , \;
{\cal J}_b[\wit{\vph}\,] = -T \int\!\D t\D \vec{r}\: \wit{\vph}^2 - \frac{\Delta_0}{2} \int \!\D\vec{r}\: \wit{\vph}_0^2 
\EEQ
with $\wit{\vph}_0=\left.\wit{\vph}\right|_{t=0}$. The dynamical functional, or action, ${\cal J}[\vph,\wit{\vph}]$ 
depends on both the order-parameter scaling operator $\vph$ and the associated response operator $\wit{\vph}$. 

Using the Euler-Lagrange equations of motion, derived from the action (\ref{17}), 
scaling and response operators are schematically characterised as follows  \cite{Henkel94}
\begin{center}\begin{tabular}{|lll|}
%\hline\noalign{\smallskip}
\noalign{\smallskip}\hline\noalign{\smallskip}
scaling operator $\vph$ :~       & scaling dimension $x$       & mass ${\cal M}>0$  \\
response operator $\wit{\vph}$ : & scaling dimension $\wit{x}$ & mass $\wit{\cal M}=-{\cal M}<0$  \\
\noalign{\smallskip}\hline
\end{tabular}\end{center}

\subsection{Bargman superselection rule, again}

If one defines the combined $(n+m)$-point functions 
\BEA
C^{(n,m)} &=& C^{(n,m)}(t_1,\ldots,t_{n+m};\vec{r}_1,\ldots,\vec{r}_{n+m}) \nonumber \\
&=&\langle \vph_1(t_1,\vec{r}_1)\ldots\vph_n(t_n,\vec{r}_n)\wit{\vph}_{n+1}(t_{n+1},\vec{r}_{n+1})\ldots\wit{\vph}_{n+m}(t_{n+m},\vec{r}_{n+m})\rangle
\nonumber
\EEA
the Bargman superselection rule (\ref{14}) can be reformulated as follows: {\it the co-variant $(n+m)$-point function $C^{(n,m)}=0$ unless $n=m$.} 
This has immediate consequences:
\begin{enumerate}
\item all co-variant correlators $C^{(n,0)}$ must vanish. 
\item only response functions $R^{(n)}=C^{(n,n)}$ can be non-vanishing. The most simple example is the two-time auto-response
$R(t,s)=C^{(1,1)}(t,s;\vec{0},\vec{0})=\langle \vph(t,\vec{0})\wit{\vph}(s;\vec{0})\rangle 
= \left.\delta\langle\vph(t;\vec{0})/\delta j(s;\vec{0})\right|_{j=0}$. 
\end{enumerate}

\subsection{Schr\"odinger-covariant response functions}

The co-variant two-time response function $R(t,s;\vec{r}_1-\vec{r}_2)=\langle \vph(t;\vec{r}_1)\wit{\vph}(s;\vec{r}_2)\rangle$, 
built from scalar scaling and response operators, obeys the conditions 
\begin{subequations}
\begin{align}
                                                                                               \left( \partial_t + \partial_s \right) R &= 0 
\label{WschX-1} \\
\left( t\partial_t + s\partial_s + \frac{r_1}{2}\partial_{r_1}+ \frac{r_2}{2}\partial_{r_2} + \frac{x}{2} + \frac{\wit{x}}{2} \right) R &= 0 
\label{WschX0}\\
                                        \left( t^2\partial_t + s^2\partial_s +tr_1\partial_{r_1}+sr_2\partial_{r_2} +x t + \wit{x} s 
                                                                  + \frac{\cal M}{2} {r_1}^2 + \frac{\wit{\cal M}}{2} {r_2}^2 \right) R &=0 
\label{WschX1}\\
                                                                                      \left( \partial_{r_1} + \partial_{r_2} \right) R &= 0 
\label{WschY-12}\\
                                                   \left( t\partial_{r_1} + s \partial_{r_2} + {\cal M}r_1 + \wit{\cal M} r_2 \right)R &= 0 
\label{WschY12} \\
                                                                                               \left( {\cal M} + \wit{\cal M} \right)R &= 0 
\label{WschM0} 
\end{align}
\end{subequations}
which follow from the Ward identities for $X_{-1}, X_0, X_1, Y_{-1/2}, Y_{1/2}, M_0$, respectively, see table~\ref{tab2}. 
Spatial rotations were not included explicitly, since for a two-point function built from scalars,
any two spatial points can be brought onto a pre-defined line, so that the problem reduces to the case $d=1$. 
Their solution follows standard lines, essentially
analogous to conformal invariance \cite{Polyakov70}. From time- and space-translation invariance (\ref{WschX-1},\ref{WschY-12}), 
it follows $R=R(\tau,r)$, with $\tau=t-s$ and $r=r_1-r_2$. As discussed above, Galilei-invariance (\ref{WschY12}) produces the Bargman superselection
rule ${\cal M} + \wit{\cal M}=0$, in agreement with (\ref{WschM0}). Then (\ref{WschX0}) and (\ref{WschY12}) lead to the equations
\BEQ
\left( \tau\partial_{\tau} + \demi r\partial_r +\demi(x+\wit{x}) \right)R=0 \;\; , \;\;
\left( \tau\partial_r + {\cal M} r\right)R =0
\EEQ
whereas the condition (\ref{WschX1}) can be simplified, by repeated application of (\ref{WschX0},\ref{WschY12}) to the condition
\BEQ
\tau r \left( x -\wit{x}\right) R = 0
\EEQ
which hence produces the constraint $x=\wit{x}$. The final form can be found from the scaling ansatz $R=\tau^{-(x+\wit{x})/2} f(r^2/\tau)$, 
in $d$ spatial dimensions \cite{Henkel94,Henkel10}
\BEQ \label{21}
R(t,s;\vec{r}) = \langle \vph(t,\vec{r})\wit{\vph}(s,\vec{0})\rangle = 
\delta_{x,\wit{x}}\:\delta({\cal M}+\wit{\cal M})\: r_0\: (t-s)^{-x} \exp\left[ -\frac{\cal M}{2} \frac{\vec{r}^2}{t-s}\right]
\EEQ
where $r_0$ is a normalisation constant. The constraint $x=\wit{x}$ is analogous to conformal invariance \cite{Polyakov70}. However,
the constraint in the masses and the heat-kernel form of the response are specific properties of Schr\"odinger-covariance. 

The Schr\"odinger-covariant three-point response function is found similarly \cite{Henkel94}:
\BEA
\lefteqn{
\hspace{-1.8truecm}\langle \vph_1(t_1,\vec{r}_1)\vph_2(t_2,\vec{r}_2)\wit{\vph}_3(t_3,\vec{r}_3)\rangle = 
\delta({\cal M}_1+{\cal M}_2 + \wit{\cal M}_3) 
\exp\left[ -\frac{{\cal M}_1}{2}\frac{\vec{r}_{13}^2}{t_{13}}-\frac{{\cal M}_2}{2}\frac{\vec{r}_{23}^2}{t_{23}}\right]
}
\nonumber \\
&\times &  t_{13}^{-x_{13,2}/2} t_{23}^{-x_{23,1}/2} t_{12}^{-x_{12,3}/2}\: 
\Psi_{12,3}\left(\frac{(\vec{r}_{13} t_{23}-\vec{r}_{23} t_{13})^2}{t_{12}t_{13}t_{23}}\right)
\label{22} 
\EEA
where $t_{ab}=t_a-t_b$, $\vec{r}_{ab}=\vec{r}_a-\vec{r}_b$, $x_{ab,c}=x_a+x_b-x_c$ (replace $x_3 \mapsto \wit{x}_3$) and $\Psi_{12,3}$ is an arbitrary differentiable function. 
A similar expression exists for $\langle \vph_1 \wit{\vph}_2 \wit{\vph}_3\rangle$ \cite{Henkel10}. 

\subsection{Noisy responses and correlators} 

The results derived so far are consequences of the dynamical Schr\"odinger symmetry of the noise-less free diffusion equation. 
We now show how responses and correlators for noisy diffusion equations can be found. 

First, we use the decomposition ${\cal J}[\vph,\wit{\vph}]={\cal J}_0[\vph,\wit{\vph}]+{\cal J}_b[\wit{\vph}]$ of the dynamic action  and
define a {\em deterministic average}
\BEQ
\langle \mathscr{A} \rangle_0 := \int \mathscr{D}\vph \mathscr{D}\wit{\vph}\; \mathscr{A}[\vph,\wit{\vph}]\: e^{-{\cal J}_0[\vph,\wit{\vph}]}
\EEQ
If the deterministic action ${\cal J}_0$ is Schr\"odinger-invariant, 
the deterministic average will obey the Ward identities of the Schr\"odinger algebra. 

Second, the full average is rewritten as follows 
\BEA
\langle \mathscr{A} \rangle &=& \int \mathscr{D}\vph \mathscr{D}\wit{\vph}\; \mathscr{A}[\vph,\wit{\vph}]\: e^{-{\cal J}[\vph,\wit{\vph}]} 
\nonumber \\
&=& \int \mathscr{D}\vph \mathscr{D}\wit{\vph}\; 
\left( \mathscr{A}[\vph,\wit{\vph}] e^{-{\cal J}_b[\wit{\vph}]} \right) e^{-{\cal J}_0[\vph,\wit{\vph}]}
\nonumber \\
&=& \langle \mathscr{A} e^{-{\cal J}_b[\wit{\vph}]} \rangle_0
\EEA
Expanding the noise part of the action and applying the Bargman superselection rule will produce simple reduction formul{\ae} for
responses and correlators \cite{Picone04}. We shall illustrate the idea through three examples:\\

\noindent 
{\bf 1.} The noisy two-time response function is (for brevity, suppress spatial arguments)
\BEQ \label{25}
R(t,s) = \langle \vph(t) \wit{\vph}(s) e^{-{\cal J}_b[\wit{\vph}]} \rangle_0 
= \sum_{k=0}^{\infty} \frac{1}{k!} \langle \vph(t)\wit{\vph}(s) \left( -{\cal J}_b[\wit{\vph}]\right)^k \rangle_0 
= \langle \vph(t) \wit{\vph}(s) \rangle_0 = R_0(t,s)
\EEQ
Here, $R_0(t,s)$ is the noise-less response which was found from Schr\"odinger-invariance and given explicitly in (\ref{21}). 
On the other hand, $R(t,s)$ is the response which is found in an explicit model calculation or in an experiment. 
Because of the Bargman superselection rules, these two responses are identical. 
In other words, {\em the covariant two-time response function does not depend explicitly on the noise.}  
Certainly, this result does depend that a decomposition
of the dynamical action into a `deterministic' and a `noise' part is possible.\footnote{The precise structure of ${\cal J}_b$ need not be
exactly the one of (\ref{17}), but at least it should contain more $\wit{\vph}$'s than $\vph$'s for eq.~(\ref{25}) to hold.} 

Therefore, the two-point and three-point responses are given by eqs. (\ref{21},\ref{22}) 
for the noisy Langevin equation (\ref{1}), if only its deterministic part is Schr\"odinger-invariant. \\

\noindent 
{\bf 2.} The noisy two-time correlator is
\BEQ \label{26}
C(t,s) = \langle \vph(t) \vph(s) e^{-{\cal J}_b[\wit{\vph}]} \rangle_0 
= \sum_{k=0}^{\infty} \frac{1}{k!} \langle \vph(t)\vph(s) \left( -{\cal J}_b[\wit{\vph}]\right)^k \rangle_0
\EEQ
Here, two response operators $\wit{\vph}$ are needed in order to retain a non-vanishing result and consequently, the detailed structure
of ${\cal J}_b$ becomes essential. We shall use here the simple form (\ref{17}). For phase-ordering, where $T=0$, this gives
\BEQ \label{27}
C(t,s;\vec{r}) = \frac{\Delta_0}{2} \int_{\mathbb{R}^d} \!\D \vec{R}\: 
\langle \vph(t,\vec{r}+\vec{r}_0)\vph(s,\vec{r}_0) \wit{\vph}^2(0,\vec{R}) \rangle_0
\EEQ
while for interface growth, with $\Delta_0=0$, we have
\BEQ \label{28}
C(t,s;\vec{r}) = T \int_{\mathbb{R_+}\times\mathbb{R}^d} \!\!\!\D u\D\vec{R}\: 
\langle \vph(t,\vec{r}+\vec{r}_0)\vph(s,\vec{r}_0) \wit{\vph}^2(u,\vec{R}) \rangle_0
\EEQ
so that in  both cases, the correlator is found from an integral of a co-variant three-point response function, 
in turn given by (\ref{22}).\footnote{Notice that pure
responses $\langle \wit{\vph}_1 \ldots \wit{\vph}_n\rangle =0$, as it should be because of causality, see \cite{Taeuber14}.} \\

\noindent
{\bf 3.} The single-time correlator $C(t,\vec{r}) :=C(t,t;\vec{r})=\langle \vph(t,\vec{r})\vph(t,\vec{0})\rangle$ 
cannot be read off from (\ref{26},\ref{27},\ref{28})
by simply setting $t=s$. Rather, we must return to the three-point response (\ref{22}) and perform the limit $t_1-t_2\to 0$ more carefully. 
We set $\vph_1=\vph_2=\vph$, $\wit{\vph}_3={\wit{\vph}}^{\,2}$ such that the scaling dimensions $x_1=x_2=x$ and $\wit{x}_3=2\wit{x}$. The
Bargman superselection rule ${\cal M}_1+{\cal M}_2 +\wit{\cal M}_3=2{\cal M}-2{\cal M}=0$ is obeyed. Also let $t_1=t_2+\vep$ and $t_3=u$. 
For the scaling function, we make the ansatz $\Psi_{12,3}(A) = \Psi_0 A^{-\omega}$ and look for consistency in the $\vep\to 0 $ limit. 
This gives for $\vep\to 0$ (with $t>u$ implied) 
\BEA
\lefteqn{\langle \vph(t+\vep,\vec{r}_1)\vph(t,\vec{r}_2)\wit{\vph}^2(u,\vec{r}_3)\rangle_0 
=\Psi_0 \vep^{\omega-(x-\wit{x})}} 
\nonumber \\
&\times& (t-u)^{-2\wit{x}} \left( \vec{r}_1 - \vec{r}_2\right)^{-2\omega} 
\exp\left[ -\frac{\cal M}{2(t-u)}\left[ (\vec{r}_1-\vec{r}_3)^2 + (\vec{r}_2-\vec{r}_3)^2\right]\right]
\nonumber 
\EEA
The dependence on $\vep$ only disappears if $\omega=x-\wit{x}$. Since $\vec{r}=\vec{r}_1-\vec{r}_2$, we can now insert this in the explicit
expression for the single-time correlator. For example, for interface growth we have from (\ref{28})
\BEQ \label{29}
C(t,\vec{r}) = \frac{T \Psi_0}{(|\vec{r}|^2)^{x-\wit{x}}} \int_0^t \!\D u\: u^{-2\wit{x}} \int_{\mathbb{R}^d} \!\!\D\vec{R}\: 
\exp\left( -\frac{\cal M}{2u}\left[ (\vec{r}-\vec{R})^2 +\vec{R}^2 \right] \right)
\EEQ
We see that the scaling function depends not only on the scaling dimension $x$ of the scaling operator $\vph$, but also on the
scaling dimension $\wit{x}$ of the associated response operator $\wit{\vph}$. 

\hspace{-0.1truecm}For phase-ordering, replace $T$ by $\Delta_0/2$, set $u\mapsto t$ and drop the integration over $u$. 

\subsection{Tests of Schr\"odinger-invariance in the Edwards-Wilkinson model}

The {\em Edwards-Wilkinson equation} \cite{Edwards82} is the special case of (\ref{1}) with $\mathscr{V}[\vph]=0$. 
This is exactly solvable and one readily
obtains the exact expressions for the height response and correlators, 
in the frame where $\langle h(t,\vec{r})\rangle=0$ \cite{Roethlein06,Bustingorry07}
\begin{subequations} \label{30}
\begin{align}
R(t,s;\vec{r}) &= \left.\frac{\delta \langle h(t,\vec{r})\rangle}{\delta j(s,\vec{0})}\right|_{j=0} 
= r_0\, (t-s)^{-d/2} \exp\left[ -\frac{\cal M}{2} \frac{\vec{r}^2}{t-s}\right] \label{30R} \\
C(t,s;\vec{r}) &= \langle h(t,\vec{r}) h(s,\vec{0}) \rangle 
= \frac{c_0 T}{|\vec{r}|^{d-2}} \left[ \Gamma\left(\frac{d}{2}-1,\frac{\cal M}{2}\frac{\vec{r}^2}{t+s}\right) 
      - \Gamma\left(\frac{d}{2}-1,\frac{\cal M}{2}\frac{\vec{r}^2}{t-s}\right) \right] \label{30Cts}\\
C(t,\vec{r}) &= \bar{c}_0 T |\vec{r}|^{-d} \Gamma\left(\frac{d}{2}-1,\frac{\cal M}{4}\frac{\vec{r}^2}{t}\right) \label{30Ct} 
\end{align}
\end{subequations}
where $\Gamma(a,x)$ is an incomplete Gamma function \cite{Abra65} and $r_0, c_0, \bar{c}_0$ are known normalisation constants. 
In the $T\to 0$ limit, the correlators indeed vanish, as predicted by the Schr\"odinger-invariance of the noise-less diffusion equation. 

Are the exact expressions (\ref{30}) compatible with the predictions (\ref{22},\ref{28},\ref{29}) of Schr\"odinger-invariance~? \\

\noindent 
{\bf 1.} The exact response (\ref{30R}) is independent of the `temperature' $T$, 
as expected from the Bargman superselection rule and (\ref{25}). 
The precise forms of (\ref{30R}) and (\ref{22}) completely agree, so that we can identify $1+a=x=\wit{x}=d/2$. \\

\noindent
{\bf 2.} Turning to the single-time correlator (\ref{29}), we symmetrise the $\vec{R}$-integration through the 
shift $\vec{R} \mapsto \vec{R}+\demi\vec{r}$. Expanding the terms in the exponential, we find
\BEA
C(t,\vec{r}) &=& \frac{T\Psi_0}{(|\vec{r}|^2)^{x-\wit{x}}}  
\int_0^t \!\D u\: u^{-2\wit{x}} \int_{\mathbb{R}^d} \!\D\vec{R}\: 
\exp\left[-\frac{\cal M}{2u}\left[\left(\frac{\vec{r}}{2}-\vec{R}\right)^2 + \left(\frac{\vec{r}}{2}+\vec{R}\right)^2 \right] \right]
\nonumber \\
&=& \frac{T\Psi_0}{(|\vec{r}|^2)^{x-\wit{x}}}  
\int_0^t \!\D u\: u^{-2\wit{x}} \int_{\mathbb{R}^d} \!\D\vec{R}\: 
\exp\left[-\frac{\cal M}{4u}\, \vec{r}^2 \right]\, \exp\left[-\frac{\cal M}{u}\, \vec{R}^2 \right] 
\nonumber \\
&=& \frac{T\Psi_0}{(|\vec{r}|^2)^{x-\wit{x}}}\left(\frac{\pi}{\cal M}\right)^{d/2} 
\int_0^t \!\D u\: u^{d/2-2\wit{x}} \exp\left[-\frac{\cal M}{4}\frac{\vec{r}^2}{u}\right]
\nonumber \\
&=& T \bar{c}_0\: |\vec{r}|^{d-2x-2\wit{x}} \Gamma\left( 2\wit{x}-\frac{d}{2}-1,\frac{\cal M}{4}\frac{\vec{r}^2}{t}\right)
\EEA
which agrees with (\ref{30Ct}), with the same identifications as above for the response. \\

\noindent 
{\bf 3.} Finally, the identity of the exact the two-time correlator (\ref{30Cts}) with the prediction (\ref{28}), using the three-point
response (\ref{22}), is shown analogously \cite{Roethlein06}. \\

Therefore, with the same consistent identification of the scaling dimensions for all three  measurable quantities, 
we find full consistency with the predictions of Schr\"o\-din\-ger-in\-va\-riance for the Edwards-Wilkinson equation. 
That was the main purpose of this section: {\it to go through all steps of
the formulation and of the derivation of simple consequences of Schr\"odinger-invariance, such that tests of 
Schr\"odinger-invariance in a simple model can be followed closely.} 

These tests of Schr\"odinger-invariance do not depend on being able to produce an exact solution 
of the model under study. Indeed, a lattice realisation
of the Edwards-Wilkinson universality class is given by the {\em Family model} \cite{Family86}: it describes the heights $h_i(t)$ on the
sites $i\in\mathscr{L}$ of the (hypercubic) lattice $\mathscr{L}\subset\mathbb{Z}^d$. At each time step, a site $i$ is randomly
selected for a deposition attempt. Before depositing a new particle, all sites $j$ in the vicinity of $i$ (usually, one takes
the nearest neighbours) are considered and one looks for the site $j_{\rm min}$ of minimal height: 
$h_{j_{\rm min}}(t)\stackrel{!}{\leq} h_{j}(t)$. Then the particle is deposited at the site $j_{\rm min}$, i.e.
$h_{j_{\rm min}}(t+1)= h_{j_{\rm min}}(t)+1$, and all other $h_i(t)$ are unchanged at this time step. 
The procedure is repeated for the next time step. A coarse-graining procedure
shows that this reproduces the Edwards-Wilkinson equation \cite{Vvedensky03}. A careful simulational study of the Family model, 
for both $d=1$ and $d=2$, reproduces precisely the exact time-space behaviour of the two-time correlator (\ref{30Cts}) \cite{Roethlein06}
and thereby confirms the Schr\"odinger-invariance of the Family model. 

\subsection{Test of Schr\"odinger-invariance of the free gaussian field}

For phase-ordering, the predictions (\ref{22},\ref{29}) of Schr\"odinger-invariance can be adapted, with the result
\begin{subequations}
\begin{align}
R(t,s;\vec{r}) &= \delta_{x,\wit{x}}\: \delta({\cal M}+\wit{\cal M}\,) r_0\: (t-s)^{-x} \exp\left[-\frac{\cal M}{2} \frac{\vec{r}^2}{t-s} \right] \\
C(t;\vec{r}) &= \bar{c}_0 |\vec{r}|^{-2(x-\wit{x})} t^{d/2-2\wit{x}} \exp\left[-\frac{\cal M}{4} \frac{\vec{r}^2}{t} \right]
\end{align}
\end{subequations}
If one identifies $x=\wit{x}=d/4$, these predictions are indeed reproduced from the 
exact solution of the free gaussian field \cite{Janssen89,Newman90}. 

%
%%%%%%%%%%%%%%%%%%%%%%%%%%%%%%%%%%%%%%%%%%%%%%%%%%%%%%%%%%%%%%%%%%%%%%%%%%%%%%%%%%%%%%%%%%%%%%%%%%%%%%%%%%%%%%%%%%%%
\section{Ageing-invariance and \\ the spherical and Arcetri models} \label{sect3}
%%%%%%%%%%%%%%%%%%%%%%%%%%%%%%%%%%%%%%%%%%%%%%%%%%%%%%%%%%%%%%%%%%%%%%%%%%%%%%%%%%%%%%%%%%%%%%%%%%%%%%%%%%%%%%%%%%%%
%
The predictions (\ref{21},\ref{26},\ref{29}) are not the final word of local scale-invariance. We  now
illustrate what can happen with a more general form of the deterministic action ${\cal J}_0$ or the Schr\"odinger operator
$\cal S$. Of course, ageing-invariance is not the last word either. 

\subsection{Ageing algebra}

The Schr\"odinger algebra contains time-translations $X_{-1}=-\partial_t$ and hence can only describe the behaviour of systems
at their stationary state. The description of generic systems far from a stationary state requires that at least this generator is
dropped. We therefore define the {\em ageing algebra} $\mathfrak{age}(1)=\langle X_{0,1}, Y_{\pm 1/2}, M_0\rangle$ \cite{Henkel06a,Henkel10}. 
Moreover, it turns out that the generators $X_n$ now admit a more general form, namely 
\begin{subequations} \label{33} 
\begin{align}
X_n &= -t^{n+1}\partial_t - \frac{n+1}{2} t^n r\partial_r - \frac{x}{2}(n+1) t^n - \xi n t^{n} - \frac{n(n+1)}{4}{\cal M} t^{n-1} r^2 
\label{33X} \\
Y_m &= -t^{m+1/2}\partial_r - \left(m+\demi\right){\cal M} t^{m-1/2}r \label{33Y} \\
M_n &= -t^n {\cal M} \label{33M}
\end{align}
\end{subequations}
whereas the generators $Y_m$ and $M_n$ are not modified with respect to the Schr\"odinger algebra, eq.~(\ref{6}). The new feature is that now
{\em a scaling operator $\vph$ is characterised by two independent scaling dimensions $x=x_{\vph}$ and $\xi=\xi_{\vph}$}. We chose the
convention that the generator $X_0$ is unmodified with respect to (\ref{6X}).\footnote{If we were to add the generator $X_{-1}$ to the algebra,
the commutator $[X_1,X_{-1}]=2X_0$ would imply $\xi=0$.} We point out that $\xi$ cannot be re-absorbed into $x$ through
a change of variables. 

One readily verifies that the commutators (\ref{7}) also hold for $\mathfrak{age}(1)$. As for the Schr\"odinger algebra, 
there is an infinite-dimensional extension, which is called the {\em ageing-Virasoro algebra} 
$\mathfrak{av}(1)=\langle X_{k}, Y_{n+1/2}, M_n\rangle_{k\in\mathbb{N},n\in\mathbb{Z}}$. It is a true subalgebra of $\mathfrak{sv}(1)$. 
As for Schr\"odinger-invariance, the extension to any dimension $d>1$ is obvious. The r\^ole of the second scaling dimension $\xi$ becomes
more clear when considering a finite transformation $t = \beta(t')$ and $\vec{r} = \vec{r}' \dot{\beta}(t')^{1/2}$ and 
\BEQ \label{34}
\vph(t,\vec{r}) = \dot{\beta}(t')^{-x/2} \left( t'\frac{\D \ln \beta(t')}{\D t'}  \right)^{-\xi} 
\exp\left[ -\frac{{\cal M}{\vec{r}'}^2}{4} \frac{\dot{\beta}(t')}{\beta(t')} \right]
\vph'(t',\vec{r}')
\EEQ
where $\beta(t)$ is again an arbitrary, but non-decreasing, differentiable function of time, which also obeys the condition $\beta(0)=0$.
Eq.~(\ref{34}) replaces eq.~(\ref{8}) of Schr\"odinger-invariance, whereas the transformation (\ref{8Y}) remains unchanged. 

\subsection{Ageing-invariance of a generalised diffusion equation}

A first appreciation of the physical relevance of the new representation (\ref{33X}) comes from the form of the $\mathfrak{age}$-invariant
Schr\"odinger operator. This operator now takes the form \cite{Niederer74,Stoimenov13}
\BEQ \label{36}
{\cal S}=2{\cal M}\partial_t - \Delta_{\vec{r}} + 2{\cal M}\left( x+\xi-\frac{d}{2}\right) t^{-1}.
\EEQ
It differs from the Schr\"odinger operator of free diffusion by the explicitly time-dependent potential term. 
The non-vanishing commutators of $\mathfrak{age}(d)$ with $\cal S$ are
\BEQ 
\left[ {\cal S}, X_0 \right] = -{\cal S} \;\; , \;\;
\left[ {\cal S}, X_1 \right] = -2t{\cal S} 
\EEQ
Generalising from section~2, we now have: {\it the space of solutions of the generalised diffusion equation ${\cal S}\vph=0$, 
with $\cal S$ given by (\ref{36}), is $\mathfrak{age}(d)$-invariant} \cite{Niederer74,Stoimenov13}. 
Note that here no condition, neither on $x$ nor on $\xi$, has to be imposed. \\

Diffusion equations $\left(\partial_t-\Delta_{\vec{r}}-V\right)\vph=0$, with time- or space-dependent potentials $V=V(t,\vec{r})$, 
have been studied intensively, and since a long time. 
For example, the dynamical symmetry algebra for an inverse-square potential $V\sim |\vec{r}|^{-2}$ is isomorphic to $\mathfrak{sch}(d)$, 
a fact already known to Jacobi, along with the case of a free particle \cite{Jacobi1842}. 
In turn, this is related to the {\em Fick-Jacobs equation}
\BD
\partial_t \vph(t,r) = \nu \frac{\partial}{\partial r} \left[ A(r) \frac{\partial}{\partial r} \frac{\vph(t,r)}{A(r)} \right] 
\ED
which describes diffusion in a rotation-symmetric channel, of cross-sectional area $A(r)$ \cite{Jacobs67}, 
with application to diffusion in biological channels or zeolites, e.g. \cite{Bressloff13}. 
If $A(r)=A_0 r^{2\mu}$, then one can map the problem onto an inverse-square potential $V(r)=V_0\mu(\mu-1)r^{-2}$ \cite{Romero14}. 
Niederer \cite{Niederer74} gave a classification of the dynamical symmetry of the diffusion equation with 
{\em any} time-space-dependent potential $V=V(t,\vec{r})$. 
Generalised representations of the ageing algebra for Schr\"odinger operators with an arbitrary time-dependent potential have been 
found recently \cite{Minic12,Henkel15}. 

We shall see below that a potential $V\sim t^{-1}$ arises naturally in certain models
of interface growth or phase-ordering.

\subsection{Non-equilibrium field-theory and Bargman selection rules}

The dynamical functional ${\cal J}[\vph,\wit{\vph}\,]={\cal J}_0[\vph,\wit{\vph}\,]+{\cal J}_b[\wit{\vph}\,]$ now takes the form
\BEA 
{\cal J}_0[\vph,\wit{\vph}\,] &=& \int\!\D t\D \vec{r}\: 
\wit{\vph}\left( \left(2{\cal M}\partial_t -\Delta_{\vec{r}} -\frac{2{\cal M}}{t}\left(x+\xi-\frac{d}{2}\right)-j\right)\vph- \mathscr{V}'[\vph]  \right)
\nonumber \\
{\cal J}_b[\wit{\vph}\,] &=& -T \int\!\D t\D \vec{r}\: \wit{\vph}^2 - \frac{\Delta_0}{2} \int \!\D\vec{r}\: \wit{\vph}_0^2 
\label{37}
\EEA
According to the new representation of $\mathfrak{age}(d)$, we have the characterisation \cite{Henkel06a}
\begin{center}\begin{tabular}{|lll|}
%\hline\noalign{\smallskip}
\noalign{\smallskip}\hline\noalign{\smallskip}
scaling operator $\vph$ :~       & scaling dimensions $x$, $\xi$             & mass ${\cal M}>0$  \\[0.14truecm]
response operator $\wit{\vph}$ : & scaling dimensions $\wit{x}$, $\wit{\xi}$ & mass $\wit{\cal M}=-{\cal M}<0$  \\
\noalign{\smallskip}\hline
\end{tabular}\end{center}
The emergence of the second, independent scaling dimension $\xi$ in non-stationary systems is a new feature, not present in dynamical
symmetries of the stationary state, such as conformal or Schr\"odinger invariance. 

The Bargman superselection rules apply as for Schr\"odinger-invariance. In particular, the average ${C}^{(n,m)}=0$ unless $n=m$. 
The dynamical symmetries of the deterministic part will therefore fix response functions. 

\subsection{Ageing-covariant response functions}

In order to find ageing-covariant response function, one might again write down the Ward identities. However, it is more simple to rewrite
the transformation (\ref{34}), generated by the $X_n$, as follows: $t = \beta(t')$, $\vec{r} = \vec{r}' \dot{\beta}(t')^{1/2}$ and 
\BEQ \label{38}
t^{-\xi}\,\vph(t,\vec{r}) = \dot{\beta}(t')^{-(x+2\xi)/2} 
\exp\left[ -\frac{{\cal M}{\vec{r}'}^2}{4} \frac{\dot{\beta}(t')}{\beta(t')} \right] \:
{t'}^{-\xi}\, \vph'(t',\vec{r}')
\EEQ
Hence, {\em if one sets $\vph(t,\vec{r})=t^{\xi}\Phi(t,\vec{r})$, 
the scaling operator $\Phi$ is Schr\"odinger-covariant, with the scaling dimension $x+2\xi$ \cite{Henkel06a}}. 
The transformations from $Y_{n+1/2}$ and $M_n$ are unchanged.\footnote{One may extend these transformations to the entire Schr\"odinger (-Virasoro) group, 
but then time-translations are generated by $X_{-1}=-\partial_t +\xi t^{-1}$ and also
change the scaling operator $\vph$. Such modifications of $X_{-1}$ also apply to more general potentials \cite{Minic12}.}

The $\mathfrak{age}$-covariant two-point response can be read from (\ref{21}):
\BEA
\lefteqn{ R(t,s;\vec{r}) = \langle t^{\xi} \Phi(t,\vec{r}) s^{\wit{\xi}} \wit{\Phi}(s,\vec{0})\rangle }  \label{39} \\
&=& \delta_{x+2\xi,\wit{x}+2\wit{\xi}}\:\delta({\cal M}+\wit{\cal M})\: r_0\: 
s^{-(x+\wit{x})/2} \left(\frac{t}{s}\right)^{\xi} \left(\frac{t}{s}-1\right)^{-(x+2\xi)}
\exp\left[-\frac{\cal M}{2}\frac{\vec{r}^2}{t-s}\right] ~~
\nonumber 
\EEA
A similar generalisation is read from (\ref{22}) for the three-point response.

\subsection{Noisy responses and correlators} 

The Bargman superselection rules are the same as those for Schr\"odinger-invariance. Therefore, eqs.~(\ref{25},\ref{27},\ref{28}) can be taken over for
ageing-invariance as well. In particular, the two-time response does not explicitly depend on the noise. Comparing with the scaling form (\ref{3},\ref{4}) 
and the explicit expression (\ref{39}), ageing-invariance predicts
\BEA
R(t,s;\vec{r}) &=& R(t,s) \exp\left[-\frac{\cal M}{2}\frac{\vec{r}^2}{t-s}\right] \nonumber \\
R(t,s) &=& r_0\: s^{-1-a} \left(\frac{t}{s}\right)^{1+a'-\lambda_R/2} \left( \frac{t}{s} -1 \right)^{-1-a'}
\label{40}
\EEA
where the three independent exponents $a$, $a'$, $\lambda_R$ are related to the three independent scaling dimensions $x,\wit{x},\xi$ by
\BEQ \label{exposants}
1+a = \demi (x+\wit{x}) \;\; , \;\; 1+a'-\lambda_R/2 = \xi \;\; , \;\; 1+a' = x+2\xi
\EEQ
Notice that because of the constraint $x+2\xi=\wit{x}+2\wit{\xi}$, the difference $a'-a=\xi+\wit{\xi}$ 
measures the contribution of the second scaling dimensions. 

Noisy correlators can be derived, analogously to Schr\"odinger-invariance, from the three-point response functions, see \cite{Henkel10}
for details. 

\subsection{Spherical model of a ferromagnet and \\ Arcetri model of an interface}

The {\em spherical model} \cite{Berlin52} is a widely studied, classical model for magnetic ordering, 
which has a critical temperature $T_c>0$ for dimensions $d>2$. 
Its collective properties are distinct from mean-field behaviour for $d<4$, yet it 
remains exactly solvable in all dimensions. Its dynamical variables are no longer discrete Ising spins $s_i=\pm 1$, 
both rather continuous `spin' variables $s(t,\vec{r})\in\mathbb{R}$ which obey the
{\em spherical constraint} $\int \!\D\vec{r}\: \langle s^2(t,\vec{r})\rangle =1$. 
Its dynamics is given by the Langevin equation \cite{Ronca78}
\BEQ \label{Lang_MS}
\partial_t s(t,\vec{r}) = \Delta_{\vec{r}} s(t,\vec{r}) + \mathfrak{z}(t) s(t,\vec{r}) + \left(2T\right)^{1/2} \eta(t,\vec{r})
\EEQ
where $\eta$ is a standard white noise of unit variance and $\mathfrak{z}(t)$ is a Lagrange multiplier to ensure the spherical constraint, 
at all times. The solution is conveniently specified in terms of the function $g(t) = \exp\left[ -2\int_0^t \!\D\tau \mathfrak{z}(\tau)\right]$. 
Because of the spherical constraint, it obeys a Volterra integral equation \cite{Godreche00b}
\BEQ \label{Volterra}
g(t) = f(t) + 2T \int_0^{t} \!\D\tau\: f(t-\tau) g(\tau)
\EEQ
where $f(t) = \left(e^{-4t} I_0(4t) \right)^d$, in the case of a totally disordered initial state ($I_0$ is a modified Bessel function \cite{Abra65}). 
If $T\leq T_c$, the long-time behaviour $g(t)\sim t^{\digamma}$ is found \cite{Godreche00b}. 
For example, phase-ordering occurs at $T=0$, when $\digamma=-d/2$ 
(for $d>2$, this remains true for all temperatures $T<T_c$). Critical dynamics occurs if $T=T_c$. If $d>4$, one finds $\digamma=0$ and one is back to
a free gaussian field. If $2<d<2$, one has $\digamma=d/2-2$. Taking the logarithm and then the derivative, this implies 
\BEQ \label{44}
\mathfrak{z}(t) \stackrel{t\to\infty}{\simeq} -\frac{\digamma}{2} \frac{1}{t} +{\rm o}(t^{-1})
\EEQ
for large times. Hence the deterministic part of the Langevin equation (\ref{Lang_MS}), for $T\leq T_c$, 
is an example of the ageing-invariant Schr\"odinger operator (\ref{36}). 
Small-time corrections to (\ref{44}) will merely generate corrections to the leading scaling behaviour.  

%%%++++++++++++++++++++++++++++++++++++++++++++++++++++++++++++++++++++++++++++++++++++++++++
\begin{figure}[tb]
\resizebox{0.68\columnwidth}{!}{%
  \includegraphics{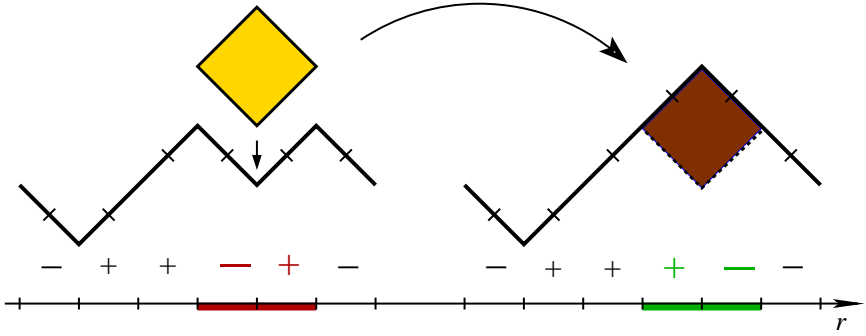} }
\caption[fig2]{Interface with the {\sc rsos} constraint $h_{i+1/2}-h_{i-1/2}=\pm 1$. A particle (left, golden) can be adsorbed (right, brown)
if the {\sc rsos} condition is still satisfied after the adsorption. The evolution of the corresponding local
slopes $u_i=h_{i+1/2}-h_{i-1/2}$ is also indicated. \label{fig2}}       
\end{figure}
%%%++++++++++++++++++++++++++++++++++++++++++++++++++++++++++++++++++++++++++++++++++++++++++

The {\em Arcetri model} \cite{Henkel15b} is an analogue of the spherical model, 
adapted to interface growth. Many lattice models of interface growth are specified
in terms of {\sc rsos} models \cite{Kim89}, see figure~\ref{fig2}, such the local slopes $u_i=h_{i+1/2}-h_{i-1/2}=\pm 1$. 
We relax this condition to a `spherical constraint' $\int\!\D \vec{r}\: u^2(t,\vec{r}) =d$ and write down the defining Langevin equation
\BEQ
\partial_t u(t,\vec{r}) = \Delta_{\vec{r}} u(t,\vec{r}) + \mathfrak{z}(t) u(t,\vec{r}) + \left(2T\right)^{1/2} \partial_r\eta(t,\vec{r})
\EEQ
in terms of the slopes $u=\partial_r h$. 
The Lagrange multiplier is analysed as in the spherical model. Again, we find the Volterra equation (\ref{Volterra}), but now with
$f(t) = \left(e^{-4t}I_0(4t)\right)^{d-1} e^{-4t} I_1(4t)/(4t)$, for a flat initial substrate. 
There is a critical point with $T_c>0$ for all $d>0$. For long times, we have
$g(t)\sim t^{\digamma}$: for $T<T_c$, $\digamma=-1-d/2$ and for $T=T_c$, $\digamma=d/2-1$ if $d<2$ but $\digamma=0$ for $d>2$. 
Therefore, for $d>2$, the Arcetri model at $T=T_c$ reduces to the Edwards-Wilkinson model. 
On the other hand, for $d<2$, the stationary exponents $z=2$ and $\beta=(2-d)/4$ are the same as for the Edwards-Wilkinson model, but the
non-stationary exponents $\lambda_C=\lambda_R=3d/2-1$ are distinct from $\lambda_C^{\rm\tiny EW} = \lambda_R^{\rm\tiny EW}=d$. 
This is an elementary example to illustrate the independence of $\lambda_C,\lambda_R$ from the stationary exponents, 
predicted long ago from field-theory \cite{Janssen89,Calabrese05,Taeuber14}. 
For $T<T_c$, eq.~(\ref{44}) applies again and the Arcetri model is an example of ageing-invariant interface growth. 

\subsection{Tests of ageing-invariance} 

Finally, we compare the prediction (\ref{40}) of ageing invariance with the exact solutions of the time-space response $R(t,s;\vec{r})$ 
of the spherical and Arcetri models. 
We find perfect agreement and extract the four scaling dimensions, as well as the three phenomenological exponents $a,a',\lambda_R$. 
This is listed in table~\ref{tab3}, where we add as well the corresponding result 
of the magnetic response of the $1D$ Ising model with Glauber dynamics, at $T=0$ \cite{Glauber63,Godreche00a} and also
those of the gaussian theory of phase-ordering \cite{Mazenko04}, also known as Ohta-Jasnow-Kawasaki ({\sc ojk}) approximation. 
Several comments are in order: 
\begin{enumerate}
\item For both the critical spherical and Arcetri models and above their upper critical dimension $d>d^*$, simple Schr\"odinger-invariance is
enough to reproduce the autoresponse. On the other hand, if fluctuation effects do become important, ageing-invariance with its second
independent scaling dimension is necessary. 
\item Since the Langevin equations of the spherical and Arcetri models are linear, the potentials $V(t)$ in the equations of motion of
$\vph$ and $\wit{\vph}$ differ by a sign. Hence $\xi+\wit{\xi}=0$ in these two models, which implies $a=a'$. 
\item The examples of the $1D$ Glauber-Ising model at $T=0$, and of the {\sc ojk}-gaussian theory,  
shows that although the equation of motion of the field $\vph$ is not linear, it should still
transform under the representation (\ref{33}) of the ageing algebra, but with $\xi+\wit{\xi}\ne 0$. Then, indeed, $a\ne a'$.  
\item The analysis of extended scaling symmetries of any more general model must begin with the identification of a `deterministic part' in the
Langevin equation and a construction of its symmetry algebra. The predictions (\ref{21},\ref{40}) are valid for certain Langevin equations only
and cannot always be applied to any other model in a straightforward manner. 
\end{enumerate}

%%%==========================================================================================
\begin{table}[tb]
\caption[tab3]{Scaling dimensions and exponents of the ageing of the autoresponse $R(t,s)$ in some exactly solved models, with $z=2$. \label{tab3}}
\begin{tabular}{lll|lllllll}  \hline\noalign{\smallskip}
model     & \multicolumn{2}{c|}{conditions} & ~$x$ & ~$\xi$   & ~$\wit{x}$ & ~$\wit{\xi}$ & ~$a$    & ~$a'$   & ~$\lambda_R$ \\
\noalign{\smallskip}\hline\noalign{\smallskip}
spherical & $T<T_c$ &                     & $0$   & $d/4$     & $d$        & $-d/4$       & $d/2-1$ & $d/2-1$ & $d/2$ \\
          & $T=T_c$ & $d<4$               & $d-2$ & $1-d/4$   & $2$        & $-1+d/4$     & $d/2-1$ & $d/2-1$ & $3d/2-2$ \\
	  & $T=T_c$ & $d>4$               & $d/2$ & $0$       & $d/2$      & $0$          & $d/2-1$ & $d/2-1$ & $d$ \\
\noalign{\smallskip}\hline\noalign{\smallskip}
Arcetri   & $T<T_c$ &                     & $-1$  & $(d+2)/4$ & $d+1$      & $-(d+2)/4$   & $d/2-1$ & $d/2-1$ & $d/2-1$ \\
          & $T=T_c$ & $d<2$               & $d-1$ & $(2-d)/4$ & $1$        & $(d-2)/4$    & $d/2-1$ & $d/2-1$ & $3d/2-1$ \\
	  & $T=T_c$ & $d>2$               & $d/2$ & $0$       & $d/2$      & $0$          & $d/2-1$ & $d/2-1$ & $d$ \\
\noalign{\smallskip}\hline\noalign{\smallskip}
Ising     & $T=0$   & $d=1$               & $1/2$ & $0$       & $3/2$      & $-1/2$       & $0$     & $-1/2$  & $1$ \\
\noalign{\smallskip}\hline\noalign{\smallskip}
\multicolumn{3}{l|}{Gaussian theory ({\sc ojk})} & $d/2$ & $0$ & $1/2$     & $(d-1)/4$    & $(d-1)/2$ & $(d-2)/2$ & $d/2$ \\
\noalign{\smallskip}\hline
\end{tabular}
\end{table}
%%%==========================================================================================

We refer to the literature for detailed accounts of tests of {\sc lsi} through the correlators \cite{Henkel10}. 
For example, in the $2D$ and $3D$ Ising models, undergoing
phase-ordering after a quench to $T\ll T_c$, the exponential spatial dependence of $R(t,s;\vec{r})$ in (\ref{21}) 
has been confirmed in detail \cite{Henkel03b}. 
Two-time correlators in Ising and Potts models undergoing phase-ordering have also been studied in great detail 
and the predictions of Schr\"odinger-invariance have been largely confirmed \cite{Henkel04,Lorenz07}.\footnote{For phase-ordering
in the $2D$ Ising model, there is a bound $\lambda_C\leq 5/4$ \cite{Fisher88}. Numerical data for $\lambda_C$ in phase-ordering 
$2D$ Ising and various Potts models usually fall slightly below this. Interestingly, the value $\lambda_C=1.25(2)$ is also found 
in the ageing of a $2D$ collapsing homopolymer model, and in agreement with the bounds
$\nu_{\rm F}d-1\leq \lambda_C\leq 2(\nu_{\rm F}d-1)$, where $\nu_F$ is the Flory exponent \cite{Majumder16}.} 
A general result of ageing-invariance is the scaling relation $\lambda_C=\lambda_R$ \cite{Picone04}, 
which had been derived before by other means \cite{Bray94}, and has been confirmed numerically many times. 
One important feature in these tests is that in analysing lattice simulations, one cannot simply use uncorrelated
initial data, as in the actions (\ref{17},\ref{37}), but some phenomenological information about the shape of the 
equal-time correlator at the onset of the scaling regime must be provided \cite{Henkel10}. 
Our new results on the equal-time correlators, outlined in section~2, might improve the situation. 

In most models, the dynamical exponent $z\ne 2$ and neither Schr\"odinger- nor ageing-invariance can be applied directly. 
However, if one restricts attention 
to the auto-response $R(t,s;\vec{0})$, then the value of the dynamical exponent only enters through the combination $\lambda_R/z$. Therefore, the
form of the autoresponse can be predicted successfully from ageing-invariance, including many instance of non-equilibrium critical dynamics. 
For example, numerical simulations of the critical Glauber-Ising model suggest $a'-a=-0.17(2)$ for $d=2$ and $a'-a=-0.022(5)$ for $d=3$
\cite{Henkel06a,Henkel10}. Studies of this kind must look very precisely into the region $t/s \gtrsim 1$ of the scaling variable, 
which requires high-precision data on huge lattices \cite{Pleimling05}. 
An  open problem is the elaboration of dynamical renormalisation-group schemes which take into account that $a'-a\ne 0$ is possible, in 
contrast to what happens at the stationary state. 

Finishing with a last outlook onto interface growth, 
the most-studied paradigm is the universality class of the Kardar-Parisi-Zhang ({\sc kpz}) equation \cite{Kardar86}
\BEQ
\partial_t h(t,\vec{r}) = \nu \Delta_{\vec{r}} h(t,\vec{r}) 
+ \frac{\mu}{2} \left( \frac{\partial h(t,\vec{r})}{\partial \vec{r}}\right)^2 +\left(2\nu T\right)^{1/2} \eta(t,\vec{r})
\EEQ
which is obtained as the continuum limit of the {\sc rsos} model \cite{Kim89} sketched in figure~\ref{fig2}. 
In the growth regime, it does undergo ageing, 
quite analogous to the Edwards-Wilkinson and Arcetri models. However, its autoresponse function cannot be described in terms of the representation
(\ref{33}) of the ageing algebra. From a phenomenological point of view, a better approximation appears to be `logarithmic representations', 
which essentially assume that the scaling operator $\vph$ acquires a `logarithmic partner' 
$\psi$ to form a doublet. Formally, one  may treat this by considering the two scaling dimensions $x,\xi$ as matrices. This leads to the form
$R(t,s) =  s^{-1-a} f_R(t/s)$, where
\BEQ \label{lnIEL}
f_R(y) = y^{-\lambda_R/z} \left(1 - \frac{1}{y}\right)^{-1-a'} 
\left[ h_0 - g_0 \ln\left( 1 - \frac{1}{y}\right) - f_0 \ln^2\left( 1 - \frac{1}{y}\right) \right]
\EEQ
and where the exponent $a'$ and the amplitudes $h_0,g_0,f_0$ must be fitted to the data \cite{Henkel10b}. 
At present, four universality classes are known
where the prediction (\ref{40}) of ageing invariance is no longer enough, but where (\ref{lnIEL}) describes the data well in the entire region where
dynamical scaling is found: {\sc kpz} for $d=1$ \cite{Henkel12} and very recently also for $d=2$ \cite{Kelling16a,Kelling16b}, 
critical directed percolation for $d=1$ \cite{Henkel10b} and the critical $2D$ Glauber-Ising model \cite{Sastre16}. 
It appears that the doublet structure only remains in the second scaling dimension $\wit{\xi}$ of the response operator. 
Taking the logarithmic terms in (\ref{lnIEL}) into account leads to improved precision 
in estimates of the exponent $\lambda_R$, which is important in order
to establish whether the equality $\lambda_C=\lambda_R$ might hold in these models, a question under active discussion \cite{Odor14,Halpin14}. 

Any substantial further progress will likely require a dynamical symmetry capable to predict the full time-space response for 
a dynamical exponent $z\ne 2$ or $z\ne 1$, which remains a difficult open problem.

%\newpage

%%%%%%%%%%%%%%%%%%%%%%%%%%%%%%%%%%%%%%%%%%%%%%%%%%%%%%%%%%%%%%%%%%%%%%%%%%%%%%%%%%%%%%%%%%%%%%%%%%%%%%%%%%%%%%%%%%

\end{document}